\documentclass[twocolumn,showpacs,preprintnumbers,amsmath,amssymb]{revtex4}

\usepackage{epsfig}

\begin{document}

\title{Effect of the exchange hole on the Gutzwiller approximation in one
  dimension}

\author{ Bal\'azs Het\'enyi, Hans Gerd Evertz, and Wolfgang von der Linden} 

\affiliation{ Institut f\"ur Theoretische Physik, Technische Universit\"at Graz, A-8010 Graz, Austria }

\begin{abstract}
  The Gutzwiller approximate solution to the Gutzwiller wavefunction yields
  exact results for the Gutzwiller wavefunction in the infinite dimensional
  limit.  Implicit in the Gutzwiller approximation is an approximate local
  form of the fermion exchange hole.  This approximate form is the same for
  all dimensions but is incorrect except in infinite dimensions.  We
  implement the correct form for the exchange hole into the Gutzwiller
  approximation.  We perform calculations on the one-dimensional Hubbard
  model at half-filling.  They indicate that the implementation of the
  exchange hole already brings the Gutzwiller approximation into very close
  quantitative agreement with the results of the full Gutzwiller
  wavefunction.  Metallicity as well as anti-ferromagnetism are recovered.
\end{abstract}

\pacs{71.10.Fd,71.30+h,71.10.Ca}


\maketitle

\section{Introduction}
\label{sec:intro}

The Hubbard model~\cite{Hubbard63,Kanamori63,Gutzwiller63,Gutzwiller65} and
its descendants have contributed greatly to our understanding of strongly
correlated systems~\cite{Imada98,Auerbach98,Fazekas99} and in particular of
the metal-insulator transition~\cite{Imada98} (MIT) exhibited by them.  Early
attempts~\cite{Gutzwiller65,Brinkman70} to explain the MIT were based on the
use of a projected wavefunction due to Gutzwiller (GWF).  The GWF is a
variational method whose starting point is a non-interacting wavefunction, in
which double occupations and consequently charge fluctuations are suppressed.

Exact solutions to the GWF are known in one~\cite{Metzner87,Metzner88} and
infinite dimensions~\cite{Metzner88,Metzner89,Metzner90}.  In one dimension
the exact solution of GWF is metallic, a conclusion which was
shown~\cite{Millis91} to be general for finite dimensions.  The exact
solution to the Hubbard model at half-filling in one-dimension~\cite{Lieb68}
is insulating for all finite values of the interaction strength.  Extended
versions of the GWF with charge fluctuations can only account for insulating
behavior when correlations between doubly occupied sites and empty sites are
incorporated (bound excitons)~\cite{Baeriswyl00,Capello05,Tahara08}.

The GWF is often treated via an approximation also due to
Gutzwiller~\cite{Gutzwiller65,Brinkman70,Fazekas99,Edegger07} (GA).  The GA
predicts a MIT~\cite{Brinkman70,Vollhardt84} between a paramagnetic metal and
a paramagnetic insulator (Brinkman-Rice transition) at half-filling, in
contradiction with the exact Gutzwiller solution.  The GA also does not
account for anti-ferromagnetic correlations properly whereas exact
diagonalizations have shown that the GWF, inspite of being based only on
projecting out double occupations, reproduces anti-ferromagnetic correlations
remarkably well~\cite{Kaplan82}.  On the other hand the GA corresponds to the
exact solution of the GWF when the number of dimensions is
infinite~\cite{Metzner88,Metzner89,Metzner90,Kotliar86}.  An improved GA has
previously been constructed by Metzner~\cite{Metzner89a} where it is shown
that self-energy corrections can restore metallicity.  Another important
study relevant here is that of van Dongen {\it et al.}\cite{Dongen89} in
which it is shown that metallicity can be recovered based on dimensional
scaling arguments, however finite orders of perturbation theory are not
sufficient to remove the Brinkman-Rice MIT.  In two dimensions metallicity
can be recovered~\cite{Gulacsi93,Gulacsi94} via a diagrammatic summation
method in which the error terms are estimated with high
accuracy~\cite{Gulacsi91}.  This has also been demonstrated
numerically~\cite{Yokoyama02}.

\begin{figure}[htp]
\vspace{1cm}
\psfig{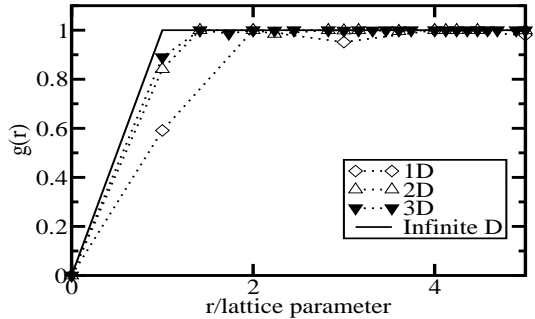}
\vspace{1cm}
\caption{Pair distribution functions of systems of non-interacting fermions
  on a lattice for different dimensions at half-filling.  The Gutzwiller
  approximation uses the pair distribution function of the inifinite limit
  for all dimensions.}
\label{fig:rdf}
\end{figure}

Interestingly, the exchange hole (pair-correlation function of particles with
parallel spins, defined as $g(r_{ij})=\langle n_i n_j \rangle$ where the
average is over the Fermi sea) for non-interacting electrons at half filling,
applied in GA is independent of the physical dimension. It corresponds to the
exact GWF result only in infinite dimensions (See Fig. \ref{fig:rdf}).  The
crucial point here is that in finite dimensions the exchange hole extends
over several lattice sites, while in GA it is restricted to the on-site term.
The central motivation of the present paper is to show that the failure of GA
is due to the over-simplified approximation of the exchange hole.

The GA consists of taking exchange into account in a combinatorial fashion.
The configurations considered obey the Pauli principle in the sense that no
two particles of the same spin can be found on the same site.  On the other
hand no other correlation effect exists between like spins in the GA.  Hence
the exchange hole is only local corresponding to the infinite dimensional
case (Fig. \ref{fig:rdf}).  It is known, however (see for example Refs.
\cite{Fazekas99} and \cite{Mahan00}) that the exchange hole in a finite
number of dimensions has a nontrivial functional form (Fig.  \ref{fig:rdf}).

The GA as well as extensions of it enjoy widespread use in a variety of
strongly correlated problems.  The Brinkman-Rice transition has been used by
Vollhardt to describe the solid-liquid phase transition in
He$^3$~\cite{Vollhardt84}.  More recent applications of the GA include
extension to the time-dependent case~\cite{Seibold01}, implementation for the
multi-band case~\cite{Bunemann00,Bunemann07}, ensembles with varying particle
number (Bardeen-Cooper-Schrieffer wavefunction)~\cite{Edegger05}, and the
calculation of matrix elements between ground and excited
states~\cite{Fukushima05}.  Variants of the approximate solution have also
been applied~\cite{Zhang88,Edegger06,Gros06} in the resonating valence bond
method~\cite{Anderson73,Fazekas74,Anderson03}, which is based on a completely
projected Gutzwiller wavefunction.

In this paper we augment the GA with an improved treatment of the exchange
hole.  Our calculation samples the exact distribution in the occupation
number representation.  The hopping term, being the expectation value of an
operator which is not diagonal in occupation number representation is
approximated in a similar fashion as in the original GA.  Implementation
consists of Monte Carlo sampling first used to calculate the exact GWF by
Yokoyama and Shiba.~\cite{Yokoyama86} It may also be possible to approximate
the distribution with the exchange hole as was done for continuous systems of
interacting fermions.~\cite{Lado67,Stevens73,Hetenyi04} For lattices a
two-site approximation which takes into account the exchange hole has been
proposed by Razafimandimby~\cite{Razafimandimby82}.

To motivate the central idea of our approach the dimensional dependence of
the exchange hole at half-filling is presented in Fig. \ref{fig:rdf}.  Due to
exchange there is an effective repulsion between nearest-neighboring ($r=1$)
particles which is strongest in the case of one dimension, and it decreases
with the number of dimensions (see also Ref.  \cite{Razafimandimby82}).
Further away from the origin ($r>1$) the functions exhibit gradually
decreasing oscillations.

Via comparison with the full GWF we show that such a procedure brings the GA
into excellent agreement with the full GWF.  The interaction energy is exact
as a function of the variational parameter, and the hopping energy also shows
close correspondence where full GWF results are available.  The
anti-ferromagnetic order parameter is also in excellent agreement between our
approach and the full GWF calculation, while the original GA significantly
underestimates antiferromagnetism.

\section{Method}
\label{sec:method}

\subsection{The Hubbard Hamiltonian and the Gutzwiller wavefunction}
\label{ssec:HMGWF}

The Hubbard Hamiltonian~\cite{Hubbard63,Kanamori63,Gutzwiller63,Gutzwiller65}
is given by
\begin{equation}
  H = -t \sum_{\langle i,j\rangle\sigma}^N c^\dagger_{i\sigma}c_{j\sigma} + U \sum_{i=1}^N n_{i\uparrow}n_{i\downarrow}.
\end{equation}
We will assume a system with $L$ lattice sites and with $N_\uparrow$ and
$N_{\downarrow}$ particles with spins up and down respectively.  For future
reference we also define the anti-ferromagnetic order parameter as
\begin{equation}
  \langle M_z^2 \rangle = \left\langle \left( \frac{1}{N}\sum_{i=1}^N
      \epsilon_i S_z(i)\right)^2 \right\rangle,
\label{eqn:afmop}
\end{equation}
where $S_z(i)$ denotes the $z$-component of the spin at site $i$,
$\epsilon_i$ is either $1$ or $-1$ depending on which sublattice site $i$
belongs to.

The variational wavefunction to the Hubbard model with which we are concerned
is the Gutzwiller wavefunction (GWF) 
\begin{equation}
|\Psi \rangle = \mbox{exp}\left(-\gamma \sum_i
n_{i\uparrow}n_{i\downarrow}\right)|FS\rangle,
\label{eqn:Psi_G}
\end{equation}
where $|FS\rangle$ indicates a Fermi sea of non-interacting fermions, the sum
in the exponential counts the number of doubly occupied sites, and $\gamma$
is a variational parameter.  For a homogeneous system the Fermi sea is
formed by filling in the plane wave states with the lowest hopping energies,
\begin{equation}
|FS\rangle = 
c^{\dagger}_{{\bf k_1}\uparrow}...c^{\dagger}_{{\bf k_{N_{\uparrow}}}\uparrow}
c^{\dagger}_{{\bf l_1}\downarrow}...c^{\dagger}_{{\bf l_{N_{\downarrow}}}\downarrow}
|0\rangle.
\label{eqn:FS_kl}
\end{equation}
Eq. (\ref{eqn:FS_kl}) can be rewritten in terms of sums over configurations
in real-space as~\cite{Fazekas99}
\begin{eqnarray}
|FS\rangle = 
L^{-\left(\frac{(N_\uparrow+N_\downarrow)}{2}\right)}
\sum_{\{{\bf g }\}}
\sum_{\{{\bf h }\}} 
\left| \begin{array}{ccc}
e^{i {\bf k_1\cdot g_1}} & \cdots & e^{i {\bf k_1\cdot g_{N_{\uparrow}}}} \\
\vdots & \ddots &  \vdots\\
e^{i {\bf k_{N_{\uparrow}}\cdot g_1}} & \cdots & e^{i {\bf k_{N_{\uparrow}}\cdot
    g_{N_{\uparrow}}}} 
\end{array} \right| \\
\times\left| \begin{array}{ccc}
e^{i {\bf l_1\cdot h_1}} & \cdots & e^{i {\bf l_1\cdot h_{N_{\downarrow}}}} \\
\vdots & \ddots &  \vdots\\
e^{i {\bf l_{N_{\downarrow}}\cdot h_1}} & \cdots & e^{i {\bf l_{N_{\downarrow}}\cdot
    h_{N_{\downarrow}}}} 
\end{array} \right| 
c^{\dagger}_{\bf g_1\uparrow}...c^{\dagger}_{\bf g_{N_{\uparrow}}\uparrow}
c^{\dagger}_{\bf h_1\downarrow}...c^{\dagger}_{\bf h_{N_{\downarrow}}\downarrow}
|0\rangle,
\nonumber 
\label{eqn:FS_gugd}
\end{eqnarray}
where ${\bf g}$ and ${\bf h}$ denote the configurations of particles with
spin up and down respectively.  These configurations are such that at most
only one particle of each spin can occupy a particular site.  To save space
we introduce the notation $\mathfrak{D}[{\bf k;g}]$ for the determinants.
The normalization of the GWF can be written
\begin{eqnarray}
\langle \Psi | \Psi \rangle &=& L^{-\left( N_\uparrow+N_\downarrow \right) }
\sum_{\{{\bf g }\}}
\sum_{\{{\bf h }\}} 
|\mathfrak{D}[{\bf k;g}]|^2 \\
& &\times|\mathfrak{D}[{\bf l;h}]|^2\mbox{exp}[-2\gamma D({\bf g},{\bf h})], \nonumber
\end{eqnarray}
where $D({\bf g},{\bf h})$ denotes the number of double
occupations for the particular configuration of up-spin and down-spin
particles ${\bf g}$ and ${\bf h}$ respectively.  One can define the
probability distribution
\begin{equation}
  P_{GWF}({\bf g},{\bf h}) = 
|\mathfrak{D}[{\bf k;g}]|^2|\mathfrak{D}[{\bf l;h}]|^2
\mbox{exp}[-2\gamma D({\bf g},{\bf h})].
\end{equation}
The exchange hole is obtained via tracing out all but two variables
corresponding to particles with parallel spin in the distribution $P_{GWF}$
with $\gamma=0$.  With the help of $P_{GWF}$ one can write expectation values
diagonal in occupation representation.  For example, the expectation value of
the number of double occupations can be written as
\begin{equation}
\label{eqn:D_GWF}
  \left\langle \sum_i n_{i\uparrow}n_{i\downarrow} \right\rangle =
  \frac{\sum_{\{{\bf g },{\bf h}\}} P_{GWF}({\bf g},{\bf h}) D({\bf g},{\bf h})}
  {\sum_{\{{\bf g },{\bf h}\}} P_{GWF}({\bf g},{\bf h})}.
\end{equation}

The hopping term is not diagonal in the coordinate representation.  For
example, if a hopping of an up-spin particle between particular sites $i$ and
$j$ is considered we have
\begin{eqnarray}
\label{eqn:hop_GWFx}
  \langle \Psi |c^\dagger_{i\uparrow}c_{j\uparrow}| \Psi \rangle = L^{-\left(N_\uparrow+N_\downarrow\right)}
  \sum_{\{{\bf g }\}}\hspace{.001cm}'\hspace{.1cm}
  \sum_{\{{\bf h }\}} 
  \mathfrak{D}^*[{\bf k;g'}_{ij\uparrow}] \times\\
   \mathfrak{D}[{\bf k;g}]|\mathfrak{D}[{\bf l;h}]|^2 \mbox{exp}[-\gamma \{D({\bf g'}_{ij\uparrow},{\bf h})+D({\bf g},{\bf h})\}]. \nonumber
\end{eqnarray}
The hopping changes the configuration from $g$ to ${g'}_{ij\uparrow}$, hence
the determinant as well as the projection term is in general altered.  In Eq.
(\ref{eqn:hop_GWFx}) ${\bf g'}_{ij\uparrow}$ denotes configurations with site
$i$ occupied and site $j$ unoccupied and ${\bf g}$ configurations with site
$i$ unoccupied and site $j$ occupied.  The prime on the sum over the up-spin
configurations indicates that only these types of configurations enter the
summation, that allow for the hopping.

The hopping can leave the number of double occupations unchanged, increase or
decrease it by one.  Hence we can rewrite Eq.  (\ref{eqn:hop_GWFx}) as
\begin{eqnarray}
\label{eqn:hop_GWF}
  \langle \Psi |c^\dagger_{i\uparrow}c_{j\uparrow}| \Psi \rangle = L^{-(N_\uparrow+N_\downarrow)}
  \sum_{\{{\bf g }\}}\hspace{.001cm} '\hspace{.1cm}
  \sum_{\{{\bf h }\}} 
  \mathfrak{D}^*[{\bf k;g'}_{ij\uparrow}] \times\\
   \mathfrak{D}[{\bf k;g}]|\mathfrak{D}[{\bf l;h}]|^2 \mbox{exp}[-\gamma
   \{2 D({\bf g},{\bf h})+\Delta D({\bf g'}_{ij\uparrow},{\bf g};{\bf h})\} ], \nonumber
\end{eqnarray}
where $\Delta D({\bf g'}_{ij\uparrow},{\bf g};{\bf h})$ denotes the change in
double occupation when an up-spin particle hops from site $j$ to $i$ (in
other words the configuration changes from ${\bf g}$ to ${\bf
  g'}_{ij\uparrow}$).  We now define the estimators for hopping from site $j$
to $i$ as
\begin{eqnarray}
\label{eqn:hopij_est_GWF}
\chi_{GWF}^{ij\uparrow}({\bf g},{\bf h}) = -t
   \mathfrak{D}^*[{\bf k;g'}_{ij\uparrow}]/\mathfrak{D}^*[{\bf k;g}]\times \\
\nonumber
\mbox{exp}[-\gamma\Delta D({\bf g'}_{ij\uparrow},{\bf g};{\bf h}) ], \\
\nonumber
\chi_{GWF}^{ij\downarrow}({\bf g},{\bf h}) = -t
   \mathfrak{D}^*[{\bf l;h'}_{ij\downarrow}]/\mathfrak{D}^*[{\bf l;h}]\times
   \\
\nonumber
\mbox{exp}[-\gamma \Delta D({\bf g};{\bf h'}_{ij\uparrow},{\bf h}) ].
\end{eqnarray}
and use these definitions to write
\begin{equation}
\label{eqn:hopij2_est_GWF}
\tau_{GWF}^{ij\sigma}({\bf g},{\bf h})
= \left\{
\begin{array}{l l}
  \chi_{GWF}^{ij\sigma}({\bf g},{\bf h}) & \mbox{if site $j$ has a particle}\\
  & \mbox{with spin $\sigma$ and}  \\
  & \mbox{site $i$  does not.}  \\
   & \\
  0 & \mbox{otherwise.} 
\end{array} \right.
\end{equation}
Eqs. (\ref{eqn:hopij_est_GWF}) and (\ref{eqn:hopij2_est_GWF}) allow us to
write the concatenated form for the estimator for the kinetic energy as
\begin{equation}
T_{GWF}({\bf g},{\bf h}) = \sum_{ij\sigma} \tau_{GWF}^{ij\sigma}({\bf g},{\bf h}),
\end{equation}
and write the expectation value of the total energy as
\begin{equation}
\label{eqn:E}
E_{GWF} = \frac{\sum_{\{{\bf g },{\bf h}\}}
P_{GWF}({\bf g},{\bf h}) 
\{
T_{GWF}({\bf g},{\bf h}) + U D({\bf g},{\bf h})
\}
}{\sum_{\{{\bf g },{\bf h}\}}P_{GWF}({\bf g},{\bf h}) } 
\end{equation}
where the summations are now unrestricted.  A Monte Carlo procedure can be
constructed~\cite{Yokoyama86} to sample the distributions $P_{GWF}$ and
evaluate the expectation values defined in (Eq. (\ref{eqn:E})).

The expectation value of the energy for the GWF was solved exactly in one
dimension by Metzner and Vollhardt~\cite{Metzner87}.  The Fermi step in this
case is finite for all finite values of the interaction strength, hence the
Gutzwiller wavefunction was shown to be metallic in one dimension.  Millis
and Coppersmith have later generalized this conclusion to any system of
finite dimensions~\cite{Millis91}.

\subsection{The Gutzwiller approximation}
\label{ssec:GA}

In the Gutzwiller approximation the determinant factors of the probability
distribution $P_{GWF}$ and of the estimator $T_{GWF}$ (Eq. (\ref{eqn:hop_GWF}))
is replaced by configurational averages obtained from the non-interacting
system.  Our description of how this is done is based on Reference
\cite{Fazekas99}.

Considering only the up-spin channel one can write the normalization of the
Fermi sea as
\begin{equation}
  _{\uparrow}\langle FS | FS \rangle_{\uparrow} =
  L^{-N_\uparrow} \sum_{\bf g} |\mathfrak{D}[{\bf k;g}]|^2 = 1,
\label{eqn:norm_up}
\end{equation}
since the wavefunctions that enter are normalized planewaves themselves.  As
the sum in Eq. (\ref{eqn:norm_up}) is over all configurations of up-spin
particles on the lattice, such that at most one particle occupies a
particular site we can approximate each term by its average as
\begin{equation}
  |\mathfrak{D}[{\bf k;g}]|^2 \approx \langle |\mathfrak{D}[{\bf k;g}]|^2\rangle = \frac{L^{N_\uparrow}}{C^L_{N_\uparrow}},
\label{eqn:approx_norm_sl}
\end{equation}
where $C^L_{N_\uparrow}$ denotes the number of ways $N_\uparrow$ particles
can be placed on $L$ lattice sites.  The down-spin particles can be handled
similarly.  This approximation results in a simplified probability
distribution as compared to $P_{GWF}$
\begin{equation}
  P_{GA}({\bf g},{\bf h}) = \mbox{exp}[-2\gamma D({\bf g},{\bf h})],
\end{equation}
allowing the rewriting of averages for quantities diagonal in the occupation
number representation.  For example the average number of double occupations
in the GA can be written as
\begin{equation}
\label{eqn:D_GA}
  \left\langle \sum_i n_{i\uparrow}n_{i\downarrow} \right\rangle =
  \frac{\sum_{\{{\bf g },{\bf h}\}} P_{GA}({\bf g},{\bf h}) D({\bf g},{\bf h})}
  {\sum_{\{{\bf g },{\bf h}\}} P_{GA}({\bf g},{\bf h})}.
\end{equation}

Approximating the kinetic energy is complicated by the fact that the hopping is
not diagonal in the occupation number representation.  Here a configurational
average is needed for products of two determinants $\mathfrak{D}^*[{\bf
  k;g'}]\mathfrak{D}[{\bf k;g}]$ for particular hopping terms.  Considering
only the up-spin channel one can write
\begin{eqnarray}
\label{eqn:t_up1}
\mathfrak{T} &=&_{\uparrow}\langle FS |c^{\dagger}_{i\uparrow} c_{j\uparrow}|
FS \rangle_{\uparrow} \\ 
&=& L^{-N_\uparrow}\sum_{\bf  g}\hspace{.001cm}'\hspace{.1cm}
\mathfrak{D}^*[{\bf
  k;g'}_{ij\uparrow}]\mathfrak{D}[{\bf k;g}]. \nonumber
\end{eqnarray}
One can also evaluate the average hopping over the Fermi sea explicitly as,
\begin{eqnarray}
\label{eqn:t_up2}
\mathfrak{T} &=& \frac{1}{L}{\sum_{\bf k}}^* \mbox{exp}[i {\bf k\cdot (R_i
  - R_j)})],  
\end{eqnarray}
where ${\bf R_i}$ and ${\bf R_j}$ denote the pair of lattice sites involved
in the hopping, and the asterisk indicates that the sum be performed over
occupied states only.  The configurations entering the sum in Eq.
(\ref{eqn:t_up1}) are the ones with one up-spin electron on site $j$ and site
$i$ unoccupied.  Of such configurations there are $C^{L-2}_{N_\uparrow-1}$.
Hence the approximation
\begin{equation}
\mathfrak{D}^*[{\bf  k;g'}_{ij\uparrow}]\mathfrak{D}[{\bf k;g}] 
\approx
\langle \mathfrak{D}^*[{\bf  k;g'}_{ij\uparrow}]\mathfrak{D}[{\bf k;g}] \rangle
= \mathfrak{T}\frac{L^{N_\uparrow}}{C^{L-2}_{N_\uparrow-1}}
\end{equation}
can be introduced.  Using this approximation the average hopping of an
up-spin particle from site $j$ to site $i$ can be written
\begin{eqnarray}
\label{eqn:hop_GA}
  \frac{\langle \Psi |c^\dagger_{i\uparrow}c_{j\uparrow} |\Psi \rangle}
  {\langle \Psi | \Psi \rangle} 
  =  \mathfrak{T} \frac{C^L_{N_\uparrow}}{C^{L-2}_{{N_\uparrow}-1}} \times \hspace{3cm}\\
  \frac{
\sum_{\{{\bf g}\}}'
\sum_{\{{\bf h}\}}
P_{GA}({\bf g},{\bf h})\mbox{exp}[-\gamma \Delta D({\bf g'}_{ij\uparrow},{\bf g};{\bf h})]
}
{ 
\sum_{\{{\bf g}\}}
\sum_{\{{\bf h}\}}
P_{GA}({\bf g},{\bf h})
}.
 \nonumber
\end{eqnarray}

Eq. (\ref{eqn:hop_GA}) is an approximate expression for the hopping energy.
It allows to formulate an estimator for the hopping which is diagonal in the
occupation number representation.  We can then write the estimator for the
hopping from site $j$ to site $i$ (the effective operator which is averaged
in order to calculate the hopping energy) as
\begin{eqnarray}
\label{eqn:hopij_est_GA}
\chi_{GA}^{ij\uparrow}({\bf g};{\bf h})  =  \tilde{\mathfrak{T}}_{\uparrow}
\mbox{exp}[-\gamma \Delta D({\bf g'}_{ij\uparrow},{\bf g};{\bf h})],\\
\nonumber
\chi_{GA}^{ij\downarrow}({\bf g};{\bf h})  =  \tilde{\mathfrak{T}}_{\downarrow}
\mbox{exp}[-\gamma \Delta D({\bf g};{\bf h'}_{ij\downarrow},{\bf h})],
\end{eqnarray}
where 
\begin{equation}
\tilde{\mathfrak{T}}_{\sigma}=\mathfrak{T}
\frac{C^L_{N_\sigma}}{C^{L-2}_{N_\sigma-1}}.
\end{equation}  
We can use these definitions to write
\begin{equation}
\label{eqn:hopij2_est_GA}
\tau_{GA}^{ij\sigma}({\bf g},{\bf h})
= \left \{
\begin{array}{l l}
  \chi_{GA}^{ij\sigma}({\bf g},{\bf h}) & \mbox{if site $j$ has a particle}\\
   &  \mbox{with spin $\sigma$ and}\\
   &  \mbox{site $i$ does not.}\\
   &  \\
  0 & \mbox{otherwise.} \\
\end{array} \right.
\end{equation}
Eqs. (\ref{eqn:hopij_est_GA}) and (\ref{eqn:hopij2_est_GA}) allow us to write
the concatenated form for estimator of the kinetic energy again as an
unrestricted sum
\begin{equation}
T_{GA}({\bf g},{\bf h}) = \sum_{ij\sigma} \tau_{GA}^{ij\sigma}({\bf g},{\bf h}),
\end{equation}
resulting in the expectation value of the total energy as
\begin{equation}
\label{eqn:E_GA}
E_{GA} = \frac{\sum_{\{{\bf g },{\bf h}\}}
P_{GA}({\bf g},{\bf h}) 
\{
T_{GA}({\bf g},{\bf h}) + U D({\bf g},{\bf h})
\}
}{\sum_{\{{\bf g },{\bf h}\}}P_{GA}({\bf g},{\bf h}) } .
\end{equation}
Note that for the non-interacting system ($U=0$) the energy is exact by
construction (one could also define the constants
$\tilde{\mathfrak{T}}_\sigma$ from this condition).  In summary the estimator
for the hopping is the product of a scaling factor and a factor which
accounts for the change in the number of double occupations caused by the
hopping itself.

In summary the GA can be considered a two-step approximation: the exact
distribution $P_{GWF}$, which is complicated by the determinant factors, is
replaced by the simpler $P_{GA}$, and the estimator for the hopping $T_{GWF}$
is replaced by $T_{GA}$.  In the $T_{GWF}$ the quotient of determinants,
which originates from the fact that the hopping is not a diagonal operator in
the occupation number representation is replaced by an average value to
arrive at the approximation $T_{GA}$.  The approximate estimator $T_{GA}$ is
diagonal in the occupation number representation.  

In the case of the GA the average energy (Eq.  (\ref{eqn:E_GA})) and other
relevant observables can be evaluated
analytically~\cite{Gutzwiller63,Gutzwiller65,Brinkman70,Vollhardt84,Fazekas99}.
The resulting MIT, known as the Brinkman-Rice transition~\cite{Brinkman70},
is characterized by the vanishing of the expectation value of the double
occupations as well as that of the hopping energy.  The latter can be shown
to be a result of the closure of the Fermi
step~\cite{Brinkman70,Vollhardt84,Fazekas99}.

\subsection{Implementing the exchange hole}
\label{ssec:GA+X}

In the following we investigate the effect of the exchange hole on the
Gutzwiller approximation.  To this end we substitute $P_{GWF}$ for $P_{GA}$
in Eq. (\ref{eqn:E_GA}).  We also scale the constants $\mathfrak{T_{\sigma}}$
so that the hopping energy remains exact in the non-interacting limit.  We
refer to this approximation scheme as the GA-X.  For the expression of the
energy we can write
\begin{equation}
\label{eqn:E_GA-X}
E_{GA-X} = \frac{\sum_{\{{\bf g },{\bf h}\}}
P_{GWF}({\bf g},{\bf h}) 
\{
T_{GA-X}({\bf g},{\bf h}) + U D({\bf g},{\bf h})
\}
}{\sum_{\{{\bf g },{\bf h}\}}P_{GWF}({\bf g},{\bf h}) } .
\end{equation}
The estimator for the hopping $T_{GA-X}$ has the same form as $T_{GA}$, the
only difference is the scaling factor to satisfy the condition in the
non-interacting limit.

The expression for the average number of double occupations as a function of
the variational parameter $\gamma$ in our scheme is the same as in the exact
GWF case (Eq. (\ref{eqn:D_GWF})).  This does not necessarily mean that we
obtain the exact expectation value of this operator, however.  Since the
hopping is {\it approximated} as a function of the parameters $t$ and $U$,
the minimization in the parameter $\gamma$ does not guarantee that the exact
$\gamma$ is obtained.
\begin{figure}[htp]
\vspace{1cm}
\psfig{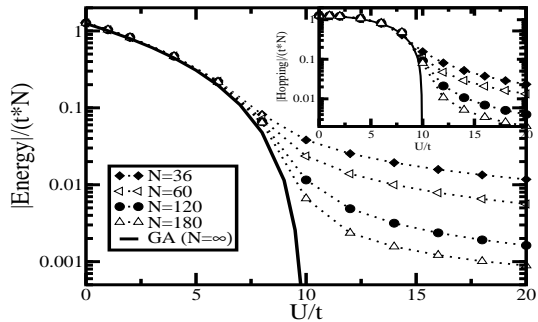}
\vspace{1cm}
\caption{Absolute value of the total energy per particle calculated in the
  Gutzwiller approximation: Monte Carlo results for various system sizes and
  analytical results for the thermodynamic limit.  The inset shows the
  absolute value of the hopping energy.}
\label{fig:nrgnox}
\end{figure}

\begin{figure}[htp]
\vspace{1cm}
\psfig{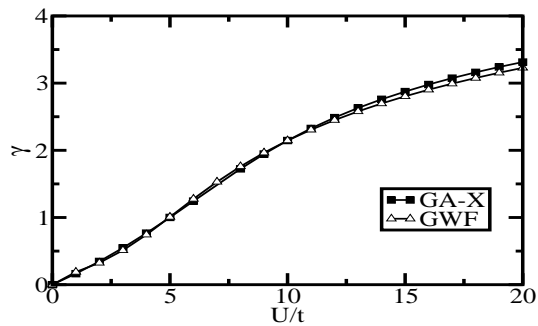}
\vspace{1cm}
\caption{Variational parameter $\gamma$ as a function of the interaction
  parameter $U$.}
\label{fig:gamma}
\end{figure}

\begin{figure}[htp]
\vspace{1cm}
\psfig{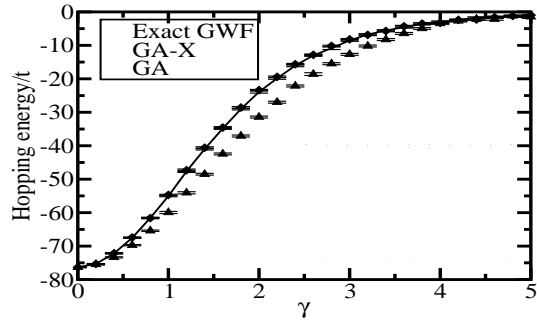}
\vspace{1cm}
\caption{Hopping energy calculated via the Monte Carlo
  sampling~\cite{Yokoyama86} of the Gutzwiller wavefunction, the Gutzwiller
  approximation with the exchange hole, and the standard Gutzwiller
  approximation for a system of $60$ sites as a function of the variational
  parameter $\gamma$.}
\label{fig:hop12}
\end{figure}

\subsection{Monte Carlo sampling}
\label{ssec:MC}

To sample the distributions described in the previous subsection we apply the
Monte Carlo method on a lattice due to Yokoyama and Shiba~\cite{Yokoyama86}.
Our system consists of a fixed number of sites $L$ and up-spin and down-spin
particles $N_\uparrow$ and $N_\downarrow$ respectively.  Our Monte Carlo
method consists of generating configurations with at most one particle of
each spin on each lattice site.  We generate the configurations using two
types of moves.  We attempt moves of particles of a particular spin to sites
without particles of that spin.  We also attempt exchange moves between sites
occupied by particles of opposite spin, and between sites which are doubly
occupied and empty.  We checked our Monte Carlo code against a full GWF
calculation based on exact diagonalization for $12$ sites and found excellent
agreement.

\section{Results}
\label{sec:Results}

In the following we apply the method described above to the one-dimensional
Hubbard model at half-filling.  In order to test our Monte Carlo program we
perform calculations for GA.  We have performed calculations with sizes up to
$180$.  The result in the thermodynamic limit is also known~\cite{Fazekas99}.
All of these calculations are based on MC runs of on the order of $10^6$
steps.  The hopping and interaction energies were calculated for an $U=1$
system.  In order to stabilize the search for the minimum, for all of the
following calculations, we fitted the calculated data points\cite{fitting}.
The results for the absolute value of the total energy are shown in Fig.
\ref{fig:nrgnox} as well as the absolute value of the kinetic energy shown in
the inset.  The Brinkman-Rice transition is clearly visible for the curve
corresponding to the thermodynamic limit.  The calculations of systems with
different sizes converge to the curve in the thermodynamic limit, in
particular they converge to zero in the region of $U$ where the Brinkman-Rice
transition predicts insulating behavior (where the energy is zero).

\begin{figure}[htp]
\vspace{1cm}
\psfig{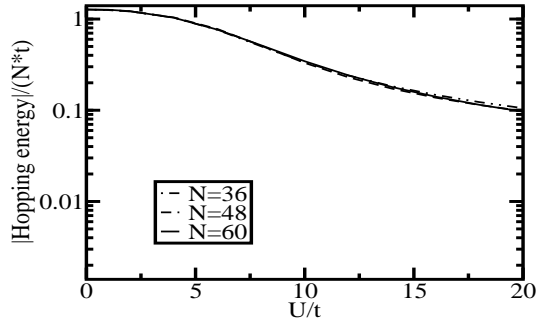}
\vspace{1cm}
\caption{Absolute value of the hopping energy calculated from the GA-X
  approximation for three different system sizes.}
\label{fig:hop_GAX}
\end{figure}

We note that often the Fermi step is taken as the indicator of the
Brinkman-Rice MIT~\cite{Fazekas99,Yokoyama86}.  To calculate the Fermi step
the density of states in momentum space is calculated.  One can argue that as
the Fermi step closes the hopping energy becomes zero, and for an open Fermi
step the hopping energy has to be finite~\cite{Fazekas99}.  Hence we can take
the hopping energy as an indicator of the MIT.  Moreover, the interaction
energy, or rather the number of double occupations, is zero as well.

\begin{figure}[htp]
\vspace{1cm}
\psfig{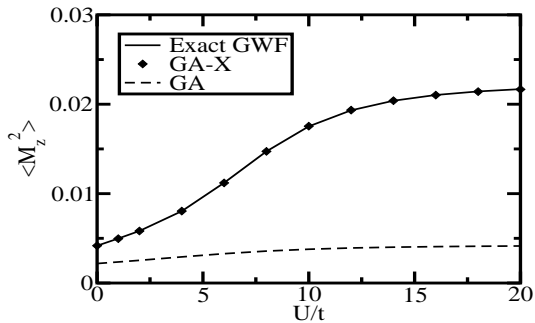}
\vspace{1cm}
\caption{Comparison of the antiferromagnetic order parameter (defined in Eq.
  (\ref{eqn:afmop})) for the exact Gutzwiller wavefunction, GA-X, and GA.}
\label{fig:afm}
\end{figure}

In Fig. \ref{fig:gamma} the variational parameter $\gamma$ is shown as a
function of the interaction parameter $U$ for the full GWF and the GA-X
calculations.  Quantitative agreement is found between the two curves.  In
Fig. \ref{fig:hop12} the hopping energies as a function of the variational
parameter $\gamma$ are compared from an exact calculation for the full GWF,
the GA-X, and the GA schemes for a system at half-filling with $60$ lattice
sites.  In the case of $60$ sites Monte Carlo sampling~\cite{Yokoyama86} was
used in all cases.  The agreement between the exact GWF and GA-X is
excellent, the exact curve essentially coincides with the GA-X results.  GWF
and GA-X differ only in the definition of the estimator for the kinetic
energy.  The fact that the kinetic energies of the two approaches essentially
coincide is indirect evidence that the GA-X approximation is metallic.  More
evidence for this conclusion is provided by comparing the hopping energy for
different system sizes, shown in Fig.  \ref{fig:hop_GAX} for sizes $36$,
$48$, and $60$.  The hopping energy, whose becoming zero indicates the
Brinkman-Rice MIT, shows negligible size-independence and does not become
zero in the range of $U$ considered (whereas the size-dependence is strong
for the GA (Fig.  \ref{fig:nrgnox})).

Since the interaction energy as a function of the variational parameter is
identical for the GWF and GA-X, our numerical evidence suggests that the GWF
and the GA-X coincide.  It is of interest to note that the GA and the GWF
coincide in the case of infinite dimensions, where the form of the exchange
hole assumed in the GA coincide with the exact exchange hole.  Our results
indicate that implementing the exact exchange hole in one dimension brings
the GA into agreement with the GWF in one dimension.

\begin{figure}[htp]
\vspace{1cm}
\psfig{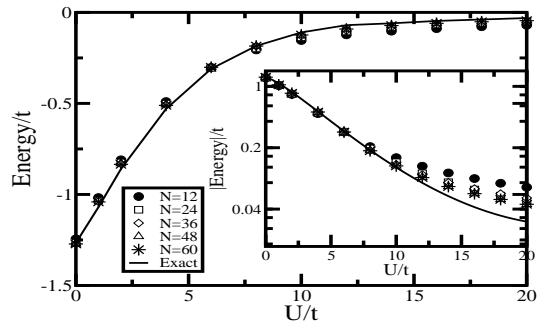}
\vspace{1cm}
\caption{Absolute value of the total energy per particle calculated via the
  GA-X: Monte Carlo results for various system sizes and the results for the
  thermodynamic limit of the full GWF (Ref. \cite{Metzner87,Metzner88}).  The
  inset (logarithmic in energy) shows that the finite size dependence of the
  energy increases with the interaction strength $U$. }
\label{fig:nrgx}
\end{figure}

\begin{figure}[htp]
\vspace{1cm}
\psfig{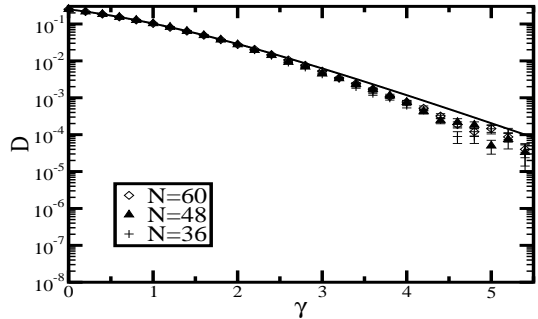}
\vspace{1cm}
\caption{Double occupation (shown on a logarithmic scale) calculated via GA-X
  and the thermodynamic limit (Ref. \cite{Metzner87,Metzner88}).  }
\label{fig:dtl}
\end{figure}

In Fig. \ref{fig:afm} we present the anti-ferromagnetic order parameter
(defined in Eq. (\ref{eqn:afmop})) for a system of $60$ lattice sites.  The
three different methods GWF, GA-X, and GA are compared.  Implementation of
the exchange hole recovers anti-ferromagnetism entirely, a results that can
be anticipated from the results on the hopping energy (Fig. \ref{fig:hop12}).
We stress that metallicity and anti-ferromagnetism are recovered together as
the exchange hole is implemented.  In Refs. \cite{Gebhard87,Gebhard88}
expressions are given for the spin-correlation functions.  Using these
expressions we have calculated the antiferromagnetic order parameter defined
in Eq. (\ref{eqn:afmop}) and we have obtained a value of $\langle M_z^2
\rangle=0.0201$ for a system of $60$ sites with $\gamma\rightarrow\infty$in
good agreement with our result of $\langle M_z^2 \rangle = 0.022\pm0.005$.
Our recovery of metallic behaviour is related to the recovery of
anti-ferromagnetic behavior.  

In Figs. \ref{fig:nrgx} and \ref{fig:dtl} we present a comparison of the
total energies and the double occupations between various system sizes.  In
the inset in Fig. \ref{fig:nrgx}, which shows the absolute value of the
energy on a logarithmic scale, the energies appear to converge to the exact
solution of the GWF in the thermodynamic limit.  The logarithmic deviations
appear to be larger at large $U$.  The average number of double occupations
as a function of the variational parameter $\gamma$ are shown in
Fig. \ref{fig:dtl}.  The exact expression for the number of double
occupations in the thermodynamic limit as a function of $\gamma$ for the GWF
wavefunction is given in Refs. \cite{Metzner87,Metzner88}.  The agreement
between the exact result and the GA-X results is goo for lower values of
$\gamma$, discrepancies appear only at large values of $\gamma$, noticible
mainly at values larger than the value that minimizes the energy at $U=20$
(see Fig. (\ref{fig:gamma})).  The discrepancy can be partly attributed to
finite size effects, which tend to oversetimate ordering, and thereby
suppress double occupations, and also the difficulty in sampling with the
standard MC method~\cite{Yokoyama86} in systems which are approaching a
critical point ($\gamma\rightarrow\infty$).

\section{Conclusions}
\label{sec:Conclusions}

We have investigated the effect of implementing the exchange hole in the
Gutzwiller approximation in one dimension for the Hubbard model at
half-filling.  The estimator used for the hopping energy was taken from the
Gutzwiller approximate solution, but the distribution of configurations was
the one corresponding to the exact solution of the Gutzwiller wavefunction,
with the exchange hole implemented.  Comparison with exact calculations based
on the Gutzwiller wavefunction were presented.

The resulting approximation is in excellent quantitative agreement with the
exact result.  The approximate hopping and the hopping of the exact solution
of the GWF are in excellent agreement.  When compared to the Gutzwiller
approximation we find that through implementing the exchange hole metallicity
and anti-ferromagnetism of the full GWF are both recovered.  Our essential
conclusion here is that by implementing the exchange hole one can account for
the anti-ferromagnetic correlations present in the exact GWF, and this
procedure results in the recovery of metallicity.

As is well known~\cite{Metzner89,Metzner90,Kotliar86} the Gutzwiller
approximation is exact in the limit of infinite dimensionality.  What our
results suggest is that implementation of the exchange hole brings the
Gutzwiller approximation into agreement with the full GWF results.  In the
future we plan to study this question in more than one finite dimensions.

From a methodological point of view our result may lead to improvements in
the future.  It is possible to construct approximate potentials for the
exchange hole~\cite{Lado67,Stevens73,Hetenyi04}.  Such an approximate
potential could then be sampled which would result in an approximate GWF, or
the exact distribution could be sampled facilitated by the approximated
potential via umbrella sampling~\cite{Torrie77}, stochastic potential
switching~\cite{Mak05} or accelerated Monte Carlo
methods\cite{Iftimie00,Hetenyi02,Bernacki04,Gelb03}.

\begin{center}
{\bf ACKNOWLEDGMENTS}
\end{center}

BH is supported by FWF (Fondsf\"orderung der wissenschaftlichen Forschung)
grant number P21240-N16.  Beneficial discussions with L. Chioncel and
E. Arrigoni are gratefully acknowledged.




\begin{references}
\bibitem{fitting} We used the following fitting procedure in determining
  continuous curves for the hopping, interaction energies, and the
  anti-ferromagnetic order parameter.  We used the software GNUPLOT version
  4.2.  For each set of data (data points for one particular system size), we
  first optimized a function of the form $f(x) = A\mbox{exp}(-Bx^C)$, then
  reoptimized using the functional form $f(x) = A\mbox{exp}(-Bx^C-Dx^E)$.  We
  found excellent agreement in all cases.
\end{references}
\end{document}